\documentstyle[12pt,epsfig,fleqn]{article}
\renewcommand{\baselinestretch}{1.2} 
\renewcommand{\arraystretch}{1.9} 
 
\newcommand{\be}{\begin{eqnarray}} 
\newcommand{\ee}{\end{eqnarray}} 
\newcommand{\nee}{\nonumber\end{eqnarray}} 
\newcommand{\nn}{\nonumber\\} 
\newcommand\sfrac[2]{{\textstyle \frac{#1}{#2}}}
\newcommand{\dgs}{d^{\gamma}{\!\scriptstyle (}s{\scriptstyle )}\,} 
\newcommand{\dzs}{d^{Z}{\!\scriptstyle (}s{\scriptstyle )}\,} 
\newcommand{\dgz}{d^{\gamma,Z}{\!\scriptstyle (}s{\scriptstyle )}\,} 
\newcommand{\mIm}{\,\mbox{\small $\Im$m\,}} 
\newcommand{\eRe}{\,\mbox{\small $\Re$e\,}} 
\newcommand{\mal}{\,\raisebox{0.3ex}{$\scriptscriptstyle \times$}\,} 
\newcommand{\Tr}{\mbox{\small Tr}}
\newcommand{\plmin}
  {\mbox{
   \mbox{\raisebox{-0.1ex}{${\scriptscriptstyle (}\!$}} 
   \mbox{\raisebox{ 0.2ex}{$\!{\displaystyle \pm}\!$}}    
   \mbox{\raisebox{-0.1ex}{$\!{\scriptscriptstyle )}$}}
  }} 
\newcommand{\gsim}{\;\raisebox{-0.3ex} 
{\small{$\stackrel{>}{\scriptstyle \sim}$}} \;} 
\newcommand{\lsim}{\;\raisebox{-0.3ex} 
{\small{$\stackrel{<}{\scriptstyle \sim}$}} \;} 
\begin{document} 
\begin{titlepage} 
\begin{flushright} 
  hep-ph/9802352 \\ 
  HEPHY--PUB 684 \\ 
  INRNE--TH--98/12 \\ 
  UWThPh--1998--2 \\
  February 1998 
\end{flushright} 
\vfill 
\begin{center} 
{\Large\bf CP violating angular asymmetries\\ 
of $b$ and $\bar{b}$ quarks in $e^{+}e^{-} \to t\bar{t}$}
\\ 
\vspace{2cm} 
{\large A. Bartl} \\ 
{\em Institut f\"ur Theoretische Physik, Universit\"at Wien, \\ 
 A-1090 Vienna, Austria} \\ 
\vspace{1cm} 
{\large E. Christova} \\ 
{\em Institute of Nuclear Research and Nuclear Energy, \\ 
 Boul. Tzarigradsko Chaussee 72, Sofia 1784, Bulgaria} \\ 
\vspace{1cm} 
{\large T. Gajdosik, W. Majerotto} \\ 
{\em Institut f\"ur Hochenergiephysik der 
 \"Osterreichischen Akademie der Wissenschaften, \\ 
 A-1050 Vienna, Austria} 
\end{center} 
\vfill

\begin{abstract}
We obtain analytical formulae for the cross section and the angular 
distributions of the $b$($\bar{b}$) quarks in the process 
$e^{+}e^{-} \to t\bar{t}$, with $t \to W^{+} b$
($\bar{t} \to W^{-}\bar{b}$) assuming CP violation in the 
\mbox{$\gamma t\bar{t}$} and \mbox{$Z t\bar{t}$} vertices. 
We present CP violating asymmetries which measure separately 
the real and imaginary parts of the 
electroweak dipole moment form factors of the top, $\dgs$ and $\dzs$. 
We give a numerical analysis of these asymmetries within the 
Minimal Supersymmetric Standard Model with complex parameters. 
They turn out to be of order $\lsim 10^{-3}$.
\end{abstract}
\end{titlepage} 
\newpage 
\setcounter{footnote}{0} 
\setcounter{page}{1} 

\section{Introduction} 
So far CP violation has been observed only in neutral 
kaon decays. An observation of {CP violation} 
in other reactions would be of crucial importance 
for a better understanding of the origin of this phenomenon. 
Top quark physics may offer new possibilities for 
{CP violating} observables at existing and future colliders. 
The reasons are both experimental and theoretical: 
Owing to its large mass the top quark decays before 
forming a hadronic bound state~\cite{Bigi}. Therefore, its 
polarization 
can be determined by measuring the distributions of its decay 
products. The polarization of the top quark is sensitive to 
{CP violation}. For isolating the truly {CP violating} effects one 
has to compare the decays of the top quark with those of the 
anti--top quark. In future $e^{+}e^{-}$ colliders, $t$ and 
$\bar{t}$ will be produced copiously, and the distributions 
of their decay products and in this way their polarizations 
can be measured in the same experiment. 

{Furthermore,} because of the large top quark mass, 
the perturbative calculations are 
more reliable and free of uncertainties of 
hadronization models. 
This allows one to obtain clear theoretical predictions for 
{CP violating} observables.

In this paper we consider the process 
\be\mbox{
\setlength{\unitlength}{1pt}\begin{picture}(160,30)(0,0)
\put(0,18){\makebox(30,12)[b]{$e^{+}e^{-}$}}
\put(33,23){\vector(1,0){24}}
\put(60,18){\makebox(8,12)[b]{$t$}}
\put(68,18){\makebox(8,12)[b]{$\bar{t}$}}
\put(64,16){\line(0,-1){3}}
\put(72,13){\oval(16,16)[bl]}
\put(72,5){\vector(1,0){12}}
\put(87,0){\makebox(30,12)[b]{$W^{+} b$}}
\end{picture}}\label{1}
\ee
and its CP--conjugate 
\be\mbox{
\setlength{\unitlength}{1pt}\begin{picture}(160,30)(0,0)
\put(0,18){\makebox(30,12)[b]{$e^{+}e^{-}$}}
\put(33,23){\vector(1,0){24}}
\put(60,18){\makebox(8,12)[b]{$t$}}
\put(68,18){\makebox(8,12)[b]{$\bar{t}$}}
\put(72,16){\line(0,-1){3}}
\put(80,13){\oval(16,16)[bl]}
\put(80,5){\vector(1,0){12}}
\put(95,0){\makebox(30,12)[b]{$W^{-} \bar{b}$}}
\end{picture}}\label{2}
\ee
in the energy range of an $e^{+}e^{-}$ Linear Collider. 
We discuss the possibility of examining {CP violation} in reactions 
(\ref{1}) and (\ref{2}) by analyzing the angular distributions of 
$b$ and $\bar{b}$ from the top quark decays. 

Previously, the effects of the electroweak dipole moment form factors 
in (\ref{1}) and (\ref{2}) were considered in \cite{soni}. 
In this paper we give analytic formulae for the differential cross 
sections of (\ref{1}) and (\ref{2}) in the c.m.system. 
We work out formulae in terms of general {CP violating} couplings 
for the angular distributions of the $b$($\bar{b}$) quark and for 
suitable defined asymmetries. It is possible to 
integrate analytically over the four--particle 
phase space. These expressions are general and model independent. 
We also take into account longitudinal polarization of 
$e^{-}$ and/or $e^{+}$.

In the Standard Model (SM) the decay $t \to b W$ has 100$\%$ 
branching ratio. We assume {CP violation} to occur in the production 
process, induced by the electric $\dgs$ and weak $\dzs$ dipole moment 
form factors of the top quark in the $\gamma t \bar{t}$ and 
$Z t \bar{t}$ vertices. In the SM $\dgs$ and $\dzs$ get non--zero 
values through the complex phase of the CKM matrix. However, they are 
at least a second order loop effect and thus almost negligible. 
{Therefore} an observation of CP non--conservation in top quark 
physics would be related to physics beyond the SM. Supersymmetric 
models and models with more than one Higgs doublet are at present 
the most favoured candidates. These models can provide new sources 
of {CP violation}~\cite{Dugan} so that 
$\dgs$ and $\dzs$ appear at one--loop level. A {complete} study of 
$\dgs$ and $\dzs$ in the Minimal Supersymmetric Standard Model (MSSM) 
with complex parameters has been performed in~\cite{dipole}. 

The previously proposed CP violating asymmetries involve 
measurements of triple product correlations~\cite{{Nacht},{Bern},
{ECMF:2},{we}}, lepton distributions, 
and other quantities~\cite{{soni},{all-the-papers},
{ECMF:1},{Rindani:1},{Rindani:2}}. 
In the present paper we show that measuring the angular distributions 
of the $b$($\bar{b}$) quarks provides an alternative method to 
determine the CP violating parameters. 
For this purpose detection of $b$ and $\bar{b}$ jets is required. 
Looking at the 
angular distributions of the {$b$($\bar{b}$) jets} coming from the 
{$t$}--decay instead of the angular distributions of the leptons 
coming from the {$W$}--decay~\cite{{Rindani:1},{Rindani:2}} has the 
advantage of a higher rate. CP violation in $t\bar{t}$ production at 
hadron colliders has been discussed in~\cite{hadron:colliders}. 

In section~2 we work out the general expressions for the differential 
cross sections of (\ref{1}) and (\ref{2}), in which the polarization 
four--vectors of $t$ and $\bar{t}$ enter explicitly. These are given 
in section~3. In section~4 we obtain 
{the differential cross section in terms of these polarization 
vectors and in section~5 we give the} 
analytic expressions in the 
c.m.system for the $\cos\theta_{b}$--distribution of the $b$ quarks 
of (\ref{1}) and of $\cos\theta_{\bar{b}}$ of the $\bar{b}$ quarks 
of process (\ref{2}), in {terms of} the real 
and imaginary parts of $\dgs$ and $\dzs$. 
Suitable angular asymmetries, sensitive to 
$\eRe \dgz$ and $\mIm \dgz$, are defined in section~6, where we also 
derive analytic formulae. In section~7 we present numerical results 
in MSSM with complex parameters, based on the 
calculations of $\dgs$ and $\dzs$ in~\cite{dipole}. 
A summary is given in section~8. 

\section{The formalism} 
In order to obtain analytic expressions for the cross sections 
of the sequential processes 
(\ref{1}) and (\ref{2}) we follow the formalism of \cite{BG}. 
{According to it we write for (\ref{1}) and (\ref{2})}: 
\be 
  d\,\sigma^{b}_{\lambda\lambda'} 
= d\,\sigma^{t}_{\lambda\lambda'} 
  \:\frac{d\,\Gamma_{\vec{t}}}{\Gamma_{t}}
  \,\frac{E_{t}}{m_{t} }
\: ,
\qquad
  d\,\sigma^{\bar{b}}_{\lambda\lambda'}
= d\,\sigma^{\bar{t}}_{\lambda\lambda'} 
  \:\frac{d\,\Gamma_{\vec{\bar{t}}}}{\Gamma_{t}}
  \,\frac{E_{\bar{t}}}{m_{t} }
\label{b:bbar} 
\ee 
Here $d\,\sigma^{t,\bar{t}}_{\lambda\lambda'}$ 
is the differential cross section for $t$ ($\bar{t}$) 
production in $e^{+}e^{-}$ annihilation, $\lambda$ and $\lambda'$ 
being the longitudinal polarization of $e^{-}$ and $e^{+}$, 
respectively. $d\,\Gamma_{\vec{t}}$ 
($d\,\Gamma_{\vec{\bar{t}}}$) is the differential 
decay rate for $t \to b W$ ($\bar{t} \to \bar{b} W$) when the 
top quark is polarized, its polarization vector $\xi$ ($\bar{\xi}$) 
determined by the former production process, $E_{t}$ ($E_{\bar{t}}$) 
is the energy of the $t$ ($\bar{t}$) quark in the c.m.system, 
$\Gamma_{t}$ is the total decay width of the top quark. 
For the differential cross section 
$d\,\sigma^{b,\bar{b}}_{\lambda,\lambda'}$ 
we obtain: 
\be & & 
  d\,\sigma^{b}_{\lambda\lambda'} 
= \sigma^{b}_{0} 
  \left\{ 1 + \alpha_{b} m_{t} \frac{(\xi p_{b})}{(p_{t} p_{b})}
  \right\} \:
  d\, \cos\theta_{t} \: d\,\Omega_{b}
\label{sigma:b-from-top}
\\ & & 
  d\,\sigma^{\bar{b}}_{\lambda\lambda'} 
= \sigma^{\bar{b}}_{0} 
  \left\{ 1 - \alpha_{b} m_{t} 
    \frac{(\bar{\xi} p_{\bar{b}})}{(p_{\bar{t}} p_{\bar{b}})}
  \right\} \:
  d\, \cos\theta_{\bar{t}} \: 
  d\,\Omega_{\bar{b}}
\label{sigma:bbar-from-top}
\ee 
We use a reference frame where the $z$--axis points into the 
direction of $\vec{\mathbf{q}}_{e}$; $\vec{\mathbf{q}}_{e}$ and 
$\vec{\mathbf{p}}_{t(\bar{t})}$ determine 
the $xz$--plane; $\cos\theta_{t(\bar{t})}$ is the scattering angle 
of $t$($\bar{t}$), and $d\,\Omega_{b(\bar{b})} 
= d\, \cos\theta_{b(\bar{b})} \: d\, \varphi_{b(\bar{b})}$. 

The coefficient $\alpha_{b}$ determines the 
sensitivity of the $b$ quark to the polarization of the top 
quark: 
\be 
  \alpha_{b} 
= \frac{m_{t}^{2} - 2 m_{\scriptscriptstyle W}^{2}}
       {m_{t}^{2} + 2 m_{\scriptscriptstyle W}^{2}}
\: .
\ee 
The sensitivity to CP non--conservation in $t$ quark production 
is determined both by the value of $\alpha_{b}$ and the 
CP violating contribution to the $t$--polarization. 
$\sigma^{b(\bar{b})}_{0}$ determines the differential SM cross 
section of (\ref{1}) and (\ref{2}) for totally unpolarized 
{decaying} top quarks with 
longitudinally polarized initial electron--positron beams: 
\be 
  \sigma^{b(\bar{b})}_{0} 
= \alpha_{em}^{2}
\:\frac{3 \beta}{2 s} 
\:\frac{\Gamma_{t\to bW}}{\Gamma_{t}} 
\:\frac{m_{t}^{2} E_{b(\bar{b})}^{2}}
       {( m_{t}^{2} - m_{\scriptscriptstyle W}^{2} )^{2}} 
\:N^{t(\bar{t})}_{\lambda\lambda'} 
\label{sigmab0} 
\ee 
where 
$E_{b(\bar{b})}$ is the energy of the $b$($\bar{b}$) quark in the 
c.m.system: 
\be 
  E_{b} 
= \frac{m_{t}^{2} - m_{\scriptscriptstyle W}^{2}}
       {\sqrt{s} \: ( 1 - \beta \cos \theta_{tb} )} 
\: , \qquad
  E_{\bar{b}} 
= \frac{m_{t}^{2} - m_{\scriptscriptstyle W}^{2}}
       {\sqrt{s} \: 
        ( 1 - \beta \cos \theta_{\,\overline{\! tb \!}\,} )} 
\: ,
\ee 
$\sqrt{s}$ is the total c.m.energy, and 
\be 
  \cos\theta_{tb} 
= \frac{(\vec{p}_{t} \!\cdot\! \vec{p}_{b})}
       {|\vec{p}_{t}|\,|\vec{p}_{b}|} 
= \sin\theta_{t} \sin\theta_{b} \cos\phi_{b} 
+ \cos\theta_{t} \cos\theta_{b}
\,. 
\ee 
We take $m_{b} = 0$, $\beta = \sqrt{1 - 4 m_{t}^{2} / s}$ is the 
velocity of the $t$ quark. 
$\Gamma_{t\to bW}$ is the partial decay width of the top quark 
for the decay $t\to bW$, and
\be 
N^{t(\bar{t})}_{\lambda\lambda'} 
= N^{t(\bar{t})}_{\lambda\lambda'}{\scriptstyle (\cos\theta_{t})}
= ( 1 + \beta^{2} \cos^{2} \theta_{t(\bar{t})} ) F_{1}
+ ( 1 - \beta^{2} ) F_{2} 
\plmin 2 \beta \cos \theta_{t(\bar{t})} \, F_{3}
\; .
\ee 
{The dependence on the beam polarizations comes through the 
functions} $F_{i}$, $i=1,2,3$, given by 
\be 
  F_{i} 
= ( 1 - \lambda \lambda' ) F_{i}^{0} 
+ ( \lambda - \lambda' ) G_{i}^{0}
\label{F} 
\ee 
where 
\be \hspace{-5mm}
F_{1}^{0} 
&=& \sfrac{4}{9} - \sfrac{4}{3} c_{V} g_{V} h_{Z} 
+ ( c_{V}^{2} + c_{A}^{2} ) ( g_{V}^{2} + g_{A}^{2} ) h_{Z}^{2}
\nn
F_{2}^{0} 
&=& \sfrac{4}{9} - \sfrac{4}{3} c_{V} g_{V} h_{Z} 
+ ( c_{V}^{2} + c_{A}^{2} ) ( g_{V}^{2} - g_{A}^{2} ) h_{Z}^{2}
\nn
F_{3}^{0} 
&=& 
- \sfrac{4}{3} c_{A} g_{A} h_{Z} 
+ 4 c_{V} c_{A} g_{V} g_{A} h_{Z}^{2}
\nn
G_{1}^{0}
&=& 
- \sfrac{4}{3} c_{A} g_{V} h_{Z} 
+ 2 c_{V} c_{A} ( g_{V}^{2} + g_{A}^{2} ) h_{Z}^{2}
\nn
G_{2}^{0}
&=& 
- \sfrac{4}{3} c_{A} g_{V} h_{Z} 
+ 2 c_{V} c_{A} ( g_{V}^{2} - g_{A}^{2} ) h_{Z}^{2}
\nn
G_{3}^{0}
&=& 
- \sfrac{4}{3} c_{V} g_{A} h_{Z} 
+ 2 ( c_{V}^{2} + c_{A}^{2} ) g_{V} g_{A} h_{Z}^{2}
\label{FG0}
\ee
The quantities 
$c_{V} = - (1/2) + 2 \sin^{2}\Theta_{\scriptscriptstyle W}$, 
$c_{A} = (1/2)$ and 
$g_{V} = (1/2) - (4/3) \sin^{2}\Theta_{\scriptscriptstyle W}$, 
$g_{A} = - (1/2)$, 
are the SM couplings of $Z$ to the electron and the top quark, 
respectively, and 
$h_{Z} = [ s / ( s - m_{Z}^{2} ) ] 
/ \sin^{2} 2 \Theta_{\scriptscriptstyle W}.$

In the SM at tree level the total cross section for the inclusive 
$b$($\bar{b}$) production process (\ref{1}) and (\ref{2}) at tree 
level is 
\be 
  \sigma^{b(\bar{b})}_{tot} 
= \frac{\pi \alpha_{em}^{2}}{s} \beta 
  \frac{\Gamma_{t \to b W}}{\Gamma_{t}}
  [ ( 3 + \beta^{2} ) F_{1} + 3 ( 1 - \beta^{2} ) F_{2} ]
= \frac{\pi \alpha_{em}^{2}}{s} \beta
  \frac{\Gamma_{t \to b W}}{\Gamma_{t}} N_{tot}
\; , 
\ee
where we have introduced the convenient notation
\be 
N_{tot} = ( 3 + \beta^{2} ) F_{1} + 3 ( 1 - \beta^{2} ) F_{2} 
\; ,
\label{NSMtot}
\ee
and the partial decay width is: 
\be 
  \Gamma_{t\to bW}
= \alpha_{em} 
  \frac{( m_{t}^{2} - m_{\scriptscriptstyle  W}^{2} )^{2} 
        ( m_{t}^{2} + 2 m_{\scriptscriptstyle  W}^{2} )}
       {16 \sin^{2} \Theta_{\scriptscriptstyle  W} \; 
        m_{t}^{3} \, m_{\scriptscriptstyle  W}^{2}}
\; . 
\ee 

\section{The polarization vector of the top quark} 
The amplitude for $e^{+}e^{-} \to t\bar{t}$, assuming 
{CP violation}, is
\be
 {\mathcal M}
&=&
  i \frac{e^{2}}{s} 
    \bar{v}{\scriptstyle (q_{\bar{e}})} \gamma_{\mu} 
    u{\scriptstyle (q_{e})} ({\mathcal V}_{\gamma})^{\mu}
- i \frac{g_{Z}^{2}}{s - m_{Z}^{2}}
    \bar{v}{\scriptstyle (q_{\bar{e}})} 
      \gamma_{\mu} ( c_{V} + c_{A} \gamma^{5} ) 
    u{\scriptstyle (q_{e})} ({\mathcal V}_{Z})^{\mu}
\label{amplitude}
\ee
where $g_{Z} = e / \sin 2 \Theta_{\scriptscriptstyle W}$. 
The quantities ${\mathcal V}_{i}$ define the 
$t \bar{t} \gamma$ and $t \bar{t} Z$ vertices\footnote{Note the
additional i in front of $\dgs$ and $\dzs$, as compared to our
previous paper~\cite{we}.}: 
\be
  ({\mathcal V}_{\gamma})^{\mu}
&=&
  \sfrac{2}{3} \gamma^{\mu} 
- i ( {\mathcal P}^{\mu} / m_{t} ) \dgs \gamma^{5}
\\
  ({\mathcal V}_{Z})^{\mu}
&=&
  \gamma^{\mu} ( g_{V} + g_{A} \gamma^{5} )
- i ( {\mathcal P}^{\mu} / m_{t} ) \dzs \gamma^{5}
\ee
Here ${\mathcal P}^{\mu} = p_{t}^{\mu} - p_{\bar{t}}^{\mu}$, and 
$\dgs$ and $\dzs$ are functions of $s$, {so} that 
$d^{\gamma}{\!\scriptstyle (0)}\,$ and 
$d^{Z}{\!\scriptstyle (m_{Z}^{2})}\,$
determine the electric and weak dipole moments of the $t$ quark. 
These dipole moments can be induced only by 
CP violating interactions and have in 
general both real and imaginary parts. 

Now we will give the expressions for the polarization four--vectors 
$\xi^{\mu}$ of the top quark and $\bar{\xi}^{\mu}$ of the antitop 
quark, depending on the electric and weak dipole moment form factors. 

As $(p_{t}\xi)=0$, in general the polarization vector $\xi^{\mu}$ 
can be decomposed along three independent four--vectors orthogonal 
to $p_{t}$: two of them, $Q_{e}^{\mu}$ and $Q_{\bar{e}}^{\mu}$ 
are in the production plane: 
\be 
Q_{e}^{\mu} 
= q_{e}^{\mu} - \frac{( p_{t} q_{e} )}{m_{t}^{2}} p_{t}^{\mu}
\; , \qquad 
Q_{\bar{e}}^{\mu}
= q_{\bar{e}}^{\mu} 
- \frac{( p_{t} q_{\bar{e}} )}{m_{t}^{2}} p_{t}^{\mu}
\; 
\ee
and the third one is normal to it: 
$\varepsilon_{\mu\alpha\beta\gamma} 
p_{t}^{\alpha} q_{e}^{\beta} q_{\bar{e}}^{\gamma}$.
Most generally, we can write: 
\be 
  \xi_{\mu} 
= P_{e}^{t} ( Q_{e} )_{\mu} 
+ P_{\bar{e}}^{t} ( Q_{\bar{e}} )_{\mu} 
+ D^{t} \varepsilon_{\mu\alpha\beta\gamma} 
        p_{t}^{\alpha} q_{e}^{\beta} q_{\bar{e}}^{\gamma}
\: .
\label{xi}
\ee
The components $P_{e(\bar{e})}^{t}$ get contributions from both 
SM and {CP violating} terms. The SM at tree level 
does not 
contribute to the normal component $\xi_{\mu}$. Thus we have: 
\be 
P_{e(\bar{e})}^{t} = P_{e(\bar{e})}^{SM} + P_{e(\bar{e})}^{CP} 
\: , \qquad
D^{t} = D^{CP}
\: . 
\label{tP} 
\ee 
The polarization four--vector is determined by the 
expression~\cite{BG}: 
\be 
  \xi_{\mu} 
= ( g_{\mu\nu} - m_{t}^{-2} p_{t\mu} p_{t\nu} )
  \Tr[ {\mathcal M} \bar{\Lambda}{\scriptstyle (p_{\bar{t}})} 
       \bar{{\mathcal M}} \Lambda{\scriptstyle (p_{t})} 
       \gamma^{\nu} \gamma^{5} ]
  \cdot 
  \Tr[ {\mathcal M} \bar{\Lambda}{\scriptstyle (p_{\bar{t}})} 
       \bar{{\mathcal M}} \Lambda{\scriptstyle (p_{t})} ]^{-1}
\ee
where ${\mathcal M}$ is the amplitude~eq.(\ref{amplitude}). 
In the c.m.system the SM contribution to $P_{e(\bar{e})}^{SM}$ is at 
tree-level 
\be 
  P_{e}^{SM}{(\theta_{t})} 
&=&
  \frac{2 m_{t}}{s} \frac{1}{N^{t}_{\lambda\lambda'}}  
  [ ( 1 - \beta \cos\theta_{t} ) ( G_{1} - G_{3} )
  + ( 1 + \beta \cos\theta_{t} ) G_{2} ]
\label{PSMe}
\\ 
  P_{\bar{e}}^{SM}{(\theta_{t})} 
&=&
- \frac{2 m_{t}}{s} \frac{1}{N^{t}_{\lambda\lambda'}} 
  [ ( 1 + \beta \cos\theta_{t} ) ( G_{1} + G_{3} )
  + ( 1 - \beta \cos\theta_{t} ) G_{2} ]
\label{PSMebar}
\ee
where $G_{i}$, $i=1,2,3$ are given by: 
\be 
  G_{i} 
= ( 1 - \lambda \lambda' ) G_{i}^{0} 
+ ( \lambda - \lambda' ) F_{i}^{0}
\label{G} 
\ee 
with $F_{i}^{0}$ and $G_{i}^{0}$ as defined in (\ref{FG0}). 
The CP violating dipole moment form factors $\dgs$ and $\dzs$
induce two types of contributions: due to their real and 
imaginary parts. The absorptive parts $\mIm \dgz$ contribute to 
$P_{e(\bar{e})}^{CP}$: 
\be 
  P_{e}^{CP}{(\theta_{t})} 
&=&
- \frac{2}{m_{t}} \frac{1}{N^{t}_{\lambda\lambda'}}
  [ ( 1 + \beta \cos\theta_{t} - \beta^{2} \sin^{2}\theta_{t} ) 
    \mIm H_{1} 
\nn & & \hspace{2cm}
  - ( \beta \cos\theta_{t} + \beta^{2} ) \mIm H_{2} ]
\label{PCPe} 
\\
  P_{\bar{e}}^{CP}{(\theta_{t})} 
&=&
- \frac{2}{m_{t}} \frac{1}{N^{t}_{\lambda\lambda'}} 
  [ ( 1 - \beta \cos\theta_{t} - \beta^{2} \sin^{2}\theta_{t} ) 
    \mIm H_{1} 
\nn & & \hspace{2cm}
  - ( \beta \cos\theta_{t} - \beta^{2} ) \mIm H_{2} ]
\; .
\label{PCPebar}
\ee
Here we have used the notation: 
\be 
  H_{i} 
= ( 1 - \lambda \lambda' ) H_{i}^{0} 
+ ( \lambda - \lambda' ) D_{i}^{0}
\label{H} 
\ee 
where 
\be 
H_{1}^{0} 
&=& ( \sfrac{2}{3} - c_{V} g_{V} h_{Z} ) \dgs
  - ( \sfrac{2}{3} c_{V} h_{Z} 
    - ( c_{V}^{2} + c_{A}^{2} ) g_{V} h_{Z}^{2} ) \dzs
\nn
H_{2}^{0} 
&=& 
  - c_{A} g_{A} h_{Z} \; \dgs
  + 2 c_{V} c_{A} g_{A} h_{Z}^{2} \; \dzs
\nn
D_{1}^{0}
&=& 
  - c_{A} g_{V} h_{Z} \; \dgs
  - ( \sfrac{2}{3} c_{A} h_{Z} 
  - 2 c_{V} c_{A} g_{V} h_{Z}^{2} ) \dzs
\nn
D_{2}^{0}
&=& 
  - c_{V} g_{A} h_{Z} \; \dgs
  + ( c_{V}^{2} + c_{A}^{2} ) g_{A} h_{Z}^{2} \; \dzs
\label{H0}
\; .
\ee
The real parts of $\dgz$ determine the CP violating contribution 
$D^{CP}$ to the normal component of the polarization vector: 
\be 
D^{CP}{(\theta_{t})} 
&=&
  \frac{8}{m_{t} s} \frac{1}{N^{t}_{\lambda\lambda'}}
  [ \eRe D_{1} + \beta \cos\theta_{t} \eRe D_{2} ]
\label{D:CP} 
\ee
Here 
\be 
  D_{i} 
= ( 1 - \lambda \lambda' ) D_{i}^{0} 
+ ( \lambda - \lambda' ) H_{i}^{0}
\label{D} 
\ee 
Note that $H_{i}^{0}$ are C--odd and P--even, while $D_{i}^{0}$ are 
C--even and P--odd functions of the coupling constants in the 
production process $e^{+}e^{-} \to t\bar{t}$. This implies that 
$H_{i}$ are C--odd and CP--odd, while $D_{i}^{0}$ are P--odd and 
CP--odd quantities. 

The polarization four--vector $\bar{\xi}$ for the anti--top is 
obtained through C--conjuga\-tion. This leads to the following 
replacements in the expressions for $\xi_{\mu}$, $F_{i}$, $G_{i}$, 
$H_{i}$, and $D_{i}$: 
\be 
p_{t} \to p_{\bar{t}} \, , \:
(2/3) e \to - (2/3) e \, , \:
g_{V} \to - g_{V} \, , \:
\dgz \to - \dgz \: .
\ee
We have: 
\be 
  \bar{\xi}_{\mu} 
= P_{e}^{\bar{t}} ( \bar{Q}_{e} )_{\mu}
+ P_{\bar{e}}^{\bar{t}} ( \bar{Q}_{\bar{e}} )_{\mu} 
+ D^{\bar{t}} \varepsilon_{\mu\alpha\beta\gamma} 
        p_{\bar{t}}^{\alpha} q_{e}^{\beta} q_{\bar{e}}^{\gamma}
\: .
\label{xibar}
\ee
where 
\be 
\bar{Q}_{e}^{\mu} 
= q_{e}^{\mu} 
- \frac{( p_{\bar{t}} q_{e} )}{m_{t}^{2}} p_{\bar{t}}^{\mu}
\; , \qquad 
\bar{Q}_{\bar{e}}^{\mu}
= q_{\bar{e}}^{\mu} 
- \frac{( p_{\bar{t}} q_{\bar{e}} )}{m_{t}^{2}} p_{\bar{t}}^{\mu}
\; .
\ee
In analogy to eq.(\ref{tP}) we define: 
\be 
P_{e(\bar{e})}^{\bar{t}} 
= \bar{P}_{e(\bar{e})}^{SM} 
+ \bar{P}_{e(\bar{e})}^{CP} 
\: , \qquad
D^{\bar{t}} = \bar{D}^{CP}
\: . 
\label{tbarP}
\ee
and obtain: 
\be 
  \bar{P}_{e}^{SM}{(\theta_{\bar{t}})} 
&=&
  \frac{2 m_{t}}{s} \frac{1}{N^{t}_{\lambda\lambda'}}  
  [ ( 1 - \beta \cos\theta_{\bar{t}} ) ( G_{1} + G_{3} )
  + ( 1 + \beta \cos\theta_{\bar{t}} ) G_{2} ]
\\
  \bar{P}_{\bar{e}}^{SM}{(\theta_{\bar{t}})} 
&=&
- \frac{2 m_{t}}{s} \frac{1}{N^{t}_{\lambda\lambda'}}  
  [ ( 1 + \beta \cos\theta_{\bar{t}} ) ( G_{1} - G_{3} )
  + ( 1 - \beta \cos\theta_{\bar{t}} ) G_{2} ]
\\ 
  \bar{P}_{e}^{CP}{(\theta_{\bar{t}})} 
&=&
- \frac{2}{m_{t}} \frac{1}{N^{t}_{\lambda\lambda'}}
  [ ( 1 + \beta \cos\theta_{\bar{t}} 
    - \beta^{2} \sin^{2}\theta_{\bar{t}} ) \mIm H_{1} 
\nn & & \hspace{2cm}
  + ( \beta \cos\theta_{\bar{t}} + \beta^{2} ) \mIm H_{2} ]
\\
  \bar{P}_{\bar{e}}^{CP}{(\theta_{\bar{t}})} 
&=&
- \frac{2}{m_{t}} \frac{1}{N^{t}_{\lambda\lambda'}}
  [ ( 1 - \beta \cos\theta_{\bar{t}}
    - \beta^{2} \sin^{2}\theta_{\bar{t}} ) \mIm H_{1} 
\nn & & \hspace{2cm}
  + ( \beta \cos\theta_{\bar{t}} - \beta^{2} ) \mIm H_{2} ]
\\
\bar{D}^{CP}{(\theta_{\bar{t}})} 
&=&
  \frac{8}{m_{t} s} \frac{1}{N^{t}_{\lambda\lambda'}}
  [ \eRe D_{1} - \beta \cos\theta_{\bar{t}} \eRe D_{2} ]
\label{Dbar:CP} 
\ee
From the explicit expressions for $\xi_{\mu}$ and $\bar{\xi}_{\mu}$ 
together with 
\be 
P^{SM}_{\pm} = P^{SM}_{e} \pm P^{SM}_{\bar{e}}
\qquad 
P^{CP}_{\pm} = P^{CP}_{e} \pm P^{CP}_{\bar{e}} 
\ee 
we obviously obtain:
\be & & 
\bar{P}_{+}^{SM}{( \theta_{\bar{t}} = \pi - \theta_{t} )}
= - P_{+}^{SM}{( \theta_{t} )}
\: , \qquad
\bar{P}_{+}^{CP}{( \theta_{\bar{t}} = \pi - \theta_{t} )}
=   P_{+}^{CP}{( \theta_{t} )}
\label{Pplus} 
\\ & &
\bar{P}_{-}^{SM}{( \theta_{\bar{t}} = \pi - \theta_{t} )}
=   P_{-}^{SM}{( \theta_{t} )}
\: , \qquad
\bar{P}_{-}^{CP}{( \theta_{\bar{t}} = \pi - \theta_{t} )}
= - P_{-}^{CP}{( \theta_{t} )}
\label{Pminus} 
\\ & &
\bar{D}^{CP}{( \theta_{\bar{t}} = \pi - \theta_{t} )}
=   D^{CP}{( \theta_{t} )}
\label{Dbar} 
\ee 

\section{The differential cross section} 
Using the explicit expressions eqs.(\ref{xi}) and (\ref{xibar}) 
for the top and the anti--top quark polarization four--vectors we 
obtain from (\ref{sigma:b-from-top}) the analytic formula for the 
cross sections of (\ref{1}) and (\ref{2}) in the c.m.system: 
\be
  d\,\sigma^{b(\bar{b})}_{\lambda\lambda'}
&=& 
  \sigma^{b(\bar{b})}_{0} {\scriptstyle (\lambda,\lambda')}
\left\{ 
  1 \plmin 
  \alpha_{b} m_{t} 
  \frac{\sqrt{s} E_{b(\bar{b})}}
       {m_{t}^{2} - m_{\scriptscriptstyle  W}^{2}} 
\left[ P^{t(\bar{t})}_{+} 
  \left( 1 
    - \frac{1 - \beta \cos \theta_{tb(\bar{tb})}}{1 - \beta^{2}} 
  \right)
\right.\right. \hspace{-15mm} 
\nn & & \hspace{15mm} 
\left.\left.
     - P^{t(\bar{t})}_{-} 
  \left( \cos\theta_{b(\bar{b})} 
     - \beta \cos \theta_{t(\bar{t})}
       \frac{1 - \beta \cos \theta_{tb(\bar{tb})}}{1 - \beta^{2}} 
  \right)
\right.\right. \hspace{-15mm} 
\nn & & \hspace{15mm} 
\left.\left.
     + D^{t(\bar{t})} \frac{s \beta}{2}
     \langle 
       \hat{\mathbf{q}}_{e} \hat{\mathbf{p}}_{t(\bar{t})} 
       \hat{\mathbf{p}}_{b(\bar{b})} 
     \rangle 
\right]
\right\}
\: d\, \cos\theta_{t(\bar{t})} \: d\,\Omega_{b(\bar{b})} 
\; ,
\label{d4sigma} 
\ee
where the triple product is defined as 
$
\langle 
  \hat{\mathbf{q}}_{e} 
  \hat{\mathbf{p}}_{t} 
  \hat{\mathbf{p}}_{b} 
\rangle 
= \hat{\mathbf{q}}_{e} \!\cdot\! 
( \hat{\mathbf{p}}_{t} \mal \hat{\mathbf{p}}_{b} )
$
with $\hat{\mathbf{q}}$, $\hat{\mathbf{p}}$ being unit three--vectors 
in the direction of the particles. $\sigma^{b(\bar{b})}_{0}$ 
is given in eq.(\ref{sigmab0}). 
We use the notation 
\be 
  P^{t}_{\pm} 
= P^{t}_{e} \pm P^{t}_{\bar{e}}
= P^{SM}_{\pm} + P^{CP}_{\pm}
\qquad
  P^{\bar{t}}_{\pm} 
= P^{\bar{t}}_{e} \pm P^{\bar{t}}_{\bar{e}}
= \bar{P}^{SM}_{\pm} + \bar{P}^{CP}_{\pm}
\; .
\ee 
$D^{t}$ and $D^{\bar{t}}$ are given by eqs.(\ref{tP}) and 
(\ref{tbarP}). 

\section{The angular distributions of $b$ and $\bar{b}$ quarks} 
Integrating (\ref{sigma:b-from-top}) and (\ref{sigma:bbar-from-top}) 
over $\cos\theta_{t}$ ($\cos\theta_{\bar{t}}$) 
and $\varphi_{b}$ ($\varphi_{\bar{b}}$) 
we obtain the $\cos\theta_{b}$ ($\cos\theta_{\bar{b}}$)--distribution 
of the $b$($\bar{b}$) quarks in the c.m.system: 
\be 
\frac{d\, \sigma^{b(\bar{b})}_{\lambda\lambda'}}
     {d\, \cos \theta_{b(\bar{b})}}
&=&
\frac{3 \pi \alpha_{em}^{2} \beta}{2 s} 
\frac{\Gamma_{t \to b W}}{\Gamma_{t}} 
\left(
    a_{0}^{b(\bar{b})}
\plmin a_{1}^{b(\bar{b})} \cos \theta_{b(\bar{b})}
  + a_{2}^{b(\bar{b})} \cos^{2} \theta_{b(\bar{b})}
\right)
\label{angle} 
\ee 
where 
\be \mbox{$\!$} \hspace{-18pt} \mbox{$\!$} & &
  a_{0}^{b(\bar{b})} = a_{0}^{SM} \plmin a_{0}^{CP}
\; , \qquad
  a_{1}^{b(\bar{b})} = a_{1}^{SM} \plmin a_{1}^{CP} 
\; , \qquad
  a_{2}^{b(\bar{b})} = a_{2}^{SM} \plmin a_{2}^{CP}
\; , \\ \mbox{$\!$} \hspace{-18pt} \mbox{$\!$} & &
  a_{0}^{SM}
\label{a0}
=
  ( 1 + \beta^{2} - b ) F_{1} + ( 1 - \beta^{2} ) F_{2}
- \alpha_{b} ( b - \beta^{2} ) G_{3}
\; , 
\\ \mbox{$\!$} \hspace{-18pt} \mbox{$\!$} & &
  a_{1}^{SM}
\label{a1}
=
  2 b F_{3}
- \alpha_{b} \left( 
      ( 1 + \beta^{2} - 2 b ) G_{1} 
    + ( 1 - \beta^{2} ) G_{2} 
  \right) 
\; , 
\\ \mbox{$\!$} \hspace{-18pt} \mbox{$\!$} & &
  a_{2}^{SM}
\label{a2}
= 
  ( 3 b - 2 \beta^{2} ) F_{1}
+ 3 \alpha_{b} ( b - \beta^{2} ) G_{3}
\; , 
\\ \mbox{$\!$} \hspace{-18pt} \mbox{$\!$} & &
  a_{0}^{CP} = - 2 \alpha_{b} b \mIm H_{1}
\; , \qquad
  a_{1}^{CP} = - 4 \alpha_{b} b \mIm H_{2}
\; , \qquad
  a_{2}^{CP} =   6 \alpha_{b} b \mIm H_{1}
\; , \qquad
\\ \mbox{$\!$} \hspace{-18pt} \mbox{$\!$} & & \quad
  b
\label{coeff-b}
=
  1 
- \sfrac{1 - \beta^{2}}{2 \beta} \ln[ \sfrac{1 + \beta}{1 - \beta} ] 
\; . 
\ee
{In ``$\plmin$'' the ``$+$'' belongs to the $b$ and 
the ``$-$'' to the $\bar{b}$.}
These formulae coincide with the analogous SM expressions 
obtained in~\cite{Segal} for the unpolarized $e^{+}e^{-}$ 
and with~\cite{me} for polarized $e^{+}e^{-}$. 
The angular distribution of the leptons {from the 
leptonic decay of $W$} is the same as that of the 
$b$ quarks because in the process $t \to bW \to b \ell \nu$ the 
$t b W$ and the $W \ell \nu$ vertices have the same Lorentz 
structure. Our 
expressions {therefore} coincide with the analogous formulae 
for the lepton angular distribution~\cite{Rindani:1}, if we replace 
$\alpha_{b}$ by $\alpha_{\ell}=1$, $\Gamma_{t \to bW}$ by 
$\Gamma_{t \to b\bar{\ell}\nu}$, and $\theta_{b}$ by $\theta_{\ell}$. 

\section{CP violating asymmetries} 
The following relation between the differential cross sections of 
processes (\ref{1}) and (\ref{2}) must hold in the case of
CP invariance: 
\be 
\frac{d\,\sigma^{b}_{\lambda\lambda'}
        {\scriptstyle (\vec{p}_{t},\vec{p}_{b})}}
     {d\,\cos \theta_{t} \: 
      d\,\Omega_{b}}
=
\frac{d\,\sigma^{\bar{b}}_{-\lambda' -\lambda}
        {\scriptstyle (\vec{p}_{\bar{t}}=-\vec{p}_{t}
                      ,\vec{p}_{\bar{b}}=-\vec{p}_{b})}}
     {d\,\cos \theta_{\bar{t}} \: 
      d\,\Omega_{\bar{b}}}
\label{sigmaCP} 
\ee
Note that in this equation (and in the following ones), the first 
lower index of $\sigma$ denotes the longitudinal polarization of the 
electron and the second one that of the positron. 

The electroweak dipole moment form factors $\dgz$ have both real and 
imaginary parts. Therefore we consider two types of observables: 
sensitive to $\eRe \dgz$ and to $\mIm \dgz$. 

First we consider observables sensitive to $\mIm \dgz$. They are 
determined by the absorptive part of the loops at the 
$t\bar{t}\gamma$ and $t\bar{t}Z$ vertices. 

\paragraph{1.} 
We shall consider two CP violating forward--backward asymmetries. Let 
$\sigma^{b(\bar{b})}_{F}{\scriptstyle (\theta_{0},\lambda,\lambda')}$ 
and 
$\sigma^{b(\bar{b})}_{B}{\scriptstyle (\theta_{0},\lambda,\lambda')}$ 
denote the number of $b$ and $\bar{b}$ quarks produced in the forward 
and backward hemispheres, respectively: 
\be 
  \sigma^{b(\bar{b})}_{F}{\scriptstyle (\theta_{0},\lambda,\lambda')} 
&=& 
  \int_{\theta_{0}}^{\pi/2} 
  \left( 
    \frac{d\, \sigma^{b(\bar{b})}_{\lambda,\lambda'}}
         {d\, \cos\theta_{b(\bar{b})}}
  \right) 
  \sin\theta_{b(\bar{b})} \: d\, \theta_{b(\bar{b})}
\: ,
\\
\sigma^{b(\bar{b})}_{B}{\scriptstyle (\theta_{0},\lambda,\lambda')}
&=& 
  \int_{\pi/2}^{\pi-\theta_{0}} 
  \left( 
    \frac{d\, \sigma^{b(\bar{b})}_{\lambda,\lambda'}}
         {d\, \cos\theta_{b(\bar{b})}}
  \right) 
  \sin\theta_{b(\bar{b})} \: d\, \theta_{b(\bar{b})}
\: .
\ee 
The standard forward--backward asymmetries of $b$ and $\bar{b}$
\be 
\mbox{A}^{b(\bar{b})}_{FB}
  {\scriptstyle (\theta_{0},\lambda,\lambda')}
=
\frac{ \sigma^{b(\bar{b})}_{F}
         {\scriptstyle (\theta_{0},\lambda,\lambda')}
     - \sigma^{b(\bar{b})}_{B}
         {\scriptstyle (\theta_{0},\lambda,\lambda')}}
     { \sigma^{b(\bar{b})}_{F}
         {\scriptstyle (\theta_{0},\lambda,\lambda')}
     + \sigma^{b(\bar{b})}_{B}
         {\scriptstyle (\theta_{0},\lambda,\lambda')}} 
\; ,
\label{Asymmetry:FB} 
\ee
define the CP violating asymmetry 
${\mathcal A}^{FB}_{\lambda\lambda'}{\scriptstyle (\theta_{0})}$
\be
  {\mathcal A}^{FB}_{\lambda\lambda'}{\scriptstyle (\theta_{0})}
= \mbox{A}_{FB}^{b}{\scriptstyle (\theta_{0},\lambda,\lambda')}
+ \mbox{A}_{FB}^{\bar{b}}
  {\scriptstyle (\theta_{0},-\lambda',-\lambda)}
\; .
\ee
From~(\ref{angle}) we obtain, keeping only the terms linear in 
\mbox{$\mIm \dgz$}, 
\be
  {\mathcal A}^{FB}_{\lambda\lambda'}{\scriptstyle (\theta_{0})}
= - 6 \alpha_{b} b \cos\theta_{0} 
\left(
  2 
  \cdot
  \frac{\mIm H_{2}}{{\mathcal N}_{tot}{\scriptstyle (\theta_{0})}} 
- 3 \sin^{2}\theta_{0} 
  \cdot
  \frac{a_{1}^{SM}}{{\mathcal N}_{tot}{\scriptstyle (\theta_{0})}} 
  \cdot
  \frac{\mIm H_{1}}{{\mathcal N}_{tot}{\scriptstyle (\theta_{0})}} 
\right)
\; ,
\label{Observable:FB} 
\ee
where ${\mathcal N}_{tot}{\scriptstyle (\theta_{0})}$ corresponds to 
the total SM cross section~(\ref{NSMtot}) with a $\theta_{0}$ cut: 
\be
  {\mathcal N}_{tot}{\scriptstyle (\theta_{0})}
= 3 a_{0}^{SM} + \cos^{2}\theta_{0} a_{2}^{SM}
= N_{tot} - \sin^{2}\theta_{0} a_{2}^{SM}
\; .
\ee
An analogous asymmetry to 
${\mathcal A}^{FB}_{\lambda\lambda'}{\scriptstyle (\theta_{0})}$ 
was considered in~\cite{{Rindani:1},{Rindani:2}}.

The difference of the number of $b$ quarks produced in 
the forward hemisphere and that of $\bar{b}$ quarks produced in the 
backward hemisphere defines our second CP violating asymmetry 
\be 
  {\mathcal A}^{A}_{\lambda\lambda'}{\scriptstyle (\theta_{0})}
=
\frac{\sigma^{b}_{F}
        {\scriptstyle (\theta_{0},\lambda,\lambda')}
     -\sigma^{\bar{b}}_{B}
        {\scriptstyle (\theta_{0},-\lambda',-\lambda)}} 
     {\sigma^{b}_{F}
        {\scriptstyle (\theta_{0},\lambda,\lambda')}
     +\sigma^{\bar{b}}_{B}
        {\scriptstyle (\theta_{0},-\lambda',-\lambda)}}
\; .
\ee 
This asymmetry is given by 
\be 
  {\mathcal A}^{A}_{\lambda\lambda'}{\scriptstyle (\theta_{0})} 
=
\frac{- 12 \alpha_{b} b 
      ( \cos\theta_{0} \mIm H_{2} + \sin^{2}\theta_{0} \mIm H_{1} )}
     { 2 {\mathcal N}_{tot}{\scriptstyle (\theta_{0})}
      + 3 \cos\theta_{0} a_{1}^{SM} 
     } 
\; . 
\label{Observable:A} 
\ee 

\paragraph{2.} In order to obtain information about $\mIm H_{1}$ we 
define a ``central'' asymmetry 
${\mathcal A}^{C}_{\lambda\lambda'}{\scriptstyle (\eta)}$ that 
measures the difference between the number of $b$ and $\bar{b}$ 
quarks in the central production 
region~\cite{{Rindani:1},{Rindani:2}}:
\be 
  {\mathcal A}^{C}_{\lambda\lambda'}{\scriptstyle (\eta)}
&=& 
\frac{ \sigma^{b}_{C}{\scriptstyle (\eta,\lambda,\lambda')}
     - \sigma^{\bar{b}}_{C}{\scriptstyle (\eta,-\lambda',-\lambda)}}
     { \sigma^{b}_{C}{\scriptstyle (\eta,\lambda,\lambda')}
     + \sigma^{\bar{b}}_{C}{\scriptstyle (\eta,-\lambda',-\lambda)}}
\ee
where
\be
  \sigma^{b(\bar{b})}_{C}{\scriptstyle (\eta,\lambda,\lambda')}
&=& 
  \int_{\eta}^{\pi-\eta} 
  \left( 
    \frac{d\, \sigma^{b(\bar{b})}_{\lambda,\lambda'}}
         {d\, \cos\theta_{b(\bar{b})}}
  \right) 
  \sin\theta_{b(\bar{b})} \: d\, \theta_{b(\bar{b})}
\: .
\ee
From~(\ref{angle}) we obtain: 
\be
  {\mathcal A}^{C}_{\lambda\lambda'}{\scriptstyle (\eta)}
&=& 
\frac{- 6 \alpha_{b} b \sin^{2}\eta \, \mIm H_{1}}
     { N_{tot} - \sin^{2}\eta \, [ ( 3 b - 2 \beta^{2} ) F_{1}
                        + 3 \alpha_{b} ( b - \beta^{2} ) G_{3} ]
     } 
\; .
\label{Observable:C} 
\ee
The magnitude of this asymmetry depends strongly on the value we 
choose for the angle $\eta$. A direct consequence of the CPT theorem 
is ${\mathcal A}^{C}_{\lambda\lambda'}{\scriptstyle (\eta=0)}=0$. 

$\mIm H_{1}$ can be measured also by the asymmetry 
${\mathcal A}^{Z}_{\lambda\lambda'}{\scriptstyle (\eta)}$:
\be
  {\mathcal A}^{Z}_{\lambda\lambda'}{\scriptstyle (\eta)}
= \mbox{A}_{Z}^{b}{\scriptstyle (\eta,\lambda,\lambda')}
- \mbox{A}_{Z}^{\bar{b}}{\scriptstyle (\eta,-\lambda',-\lambda)}
\; ,
\ee
where
\be 
  \mbox{A}^{b(\bar{b})}_{Z}{\scriptstyle (\eta,\lambda,\lambda')}
=
\frac{ \sigma^{b(\bar{b})}_{C}{\scriptstyle (\eta,\lambda,\lambda')}
     - \sigma^{b(\bar{b})}_{P}{\scriptstyle (\eta,\lambda,\lambda')}} 
     { \sigma^{b(\bar{b})}_{C}{\scriptstyle (\eta,\lambda,\lambda')}
     + \sigma^{b(\bar{b})}_{P}{\scriptstyle (\eta,\lambda,\lambda')}} 
\; .
\ee
Here $\sigma^{b(\bar{b})}_{P}$ is defined as complementary to 
$\sigma^{b(\bar{b})}_{C}$:
\be 
\sigma^{b(\bar{b})}_{P}{\scriptstyle (\eta,\lambda,\lambda')}
:= \sigma^{b(\bar{b})}_{tot}
- \sigma^{b(\bar{b})}_{C}{\scriptstyle (\eta,\lambda,\lambda')}
= 
( \int_{0}^{\eta} 
+ \int_{\pi-\eta}^{\pi} 
) \left( 
    \frac{d\, \sigma^{b(\bar{b})}_{\lambda,\lambda'}}
         {d\, \cos\theta_{b(\bar{b})}}
  \right) 
  \sin\theta_{b(\bar{b})} \: d\, \theta_{b(\bar{b})}
\: .
\ee 
For ${\mathcal A}^{Z}_{\lambda\lambda'}$ we obtain: 
\be
  {\mathcal A}^{Z}_{\lambda\lambda'}
= - 24 \alpha_{b} b \cos\eta \, \sin^{2}\eta \, \mIm H_{1} 
  / N_{tot} 
\; .
\label{Observable:Z} 
\ee
For $\cos\eta=1/\sqrt{3}$ this asymmetry is largest. 

\bigskip

The asymmetries 
${\mathcal A}^{FB}_{\lambda\lambda'}{\scriptstyle (\theta_{0})}$,
${\mathcal A}^{A}_{\lambda\lambda'}{\scriptstyle (\theta_{0})}$, 
${\mathcal A}^{C}_{\lambda\lambda'}{\scriptstyle (\eta)}$, and
${\mathcal A}^{Z}_{\lambda\lambda'}{\scriptstyle (\eta)}$ are 
different from zero if CP invariance is violated in (\ref{1}) and 
(\ref{2}). Notice that the imaginary parts of $\dgz$ enter in 
eqs. (\ref{Observable:FB}), (\ref{Observable:A}),  
(\ref{Observable:C}), and (\ref{Observable:Z}) as
${\mathcal A}^{FB}_{\lambda\lambda'}{\scriptstyle (\theta_{0})}$,
${\mathcal A}^{A}_{\lambda\lambda'}{\scriptstyle (\theta_{0})}$,
${\mathcal A}^{C}_{\lambda\lambda'}{\scriptstyle (\eta)}$
and $ {\mathcal A}^{Z}_{\lambda\lambda'}{\scriptstyle (\eta)}$
are even under time reversal transformation, in other words, the $i$ 
in front of $\dgz$ in the vertices 
${\cal V}_{\gamma}$ and ${\cal V}_{Z}$ 
is compensated by the i in front of $\mIm H_{i}$.
If $\lambda = \lambda'= 0$ these asymmetries violate 
both C and CP invariance, and in accordance with this the analytic 
expressions are proportional to the C--odd and CP--odd functions 
$H_{i}$. A measurement of these asymmetries allows one to determine 
the imaginary parts of the dipole moment form factors $\dgs$ and 
$\dzs$. 

\paragraph{3.} The real parts of $\dgz$ can be singled out by 
measuring triple product correlations~\cite{{Nacht},{ECMF:2}}. 
A suitable asymmetry is given by~\cite{we}
\be
\mbox{A}^{\! {\mathcal T}}_{b(\bar{b})}
  {\scriptstyle (\lambda,\lambda')} 
= 
( N[ \langle 
       \hat{\mathbf{q}}_{e} \hat{\mathbf{p}}_{t(\bar{t})} 
       \hat{\mathbf{p}}_{b(\bar{b})} 
     \rangle > 0] 
- N[ \langle 
       \hat{\mathbf{q}}_{e} \hat{\mathbf{p}}_{t(\bar{t})} 
       \hat{\mathbf{p}}_{b(\bar{b})} 
     \rangle < 0] ) 
/ \sigma^{b(\bar{b})}_{tot}
\: ,
\ee
where 
{\small $\langle \hat{\mathbf{q}}_{e} \hat{\mathbf{p}}_{t(\bar{t})} 
\hat{\mathbf{p}}_{b(\bar{b})} \rangle = \hat{\mathbf{q}}_{e} \!\cdot 
( \hat{\mathbf{p}}_{t(\bar{t})} \mal \hat{\mathbf{p}}_{b(\bar{b})} ) 
= \sin\theta_{t(\bar{t})} \sin\theta_{b(\bar{b})} 
  \sin\phi_{b(\bar{b})}$}.
As {\small $\sin\theta_{t(\bar{t})} ,\,\sin\theta_{b(\bar{b})}>0$}, 
{\small $N[ \langle \hat{\mathbf{q}}_{e} 
   \hat{\mathbf{p}}_{t(\bar{t})} 
   \hat{\mathbf{p}}_{b(\bar{b})} \rangle > 0 \,/ \, <0]$} 
are the number of $b$($\bar{b}$) quarks produced above/below (with 
{\small $\sin\phi_{b(\bar{b})}>0 \,/ \, <0$}) the production--plane 
{\small $\{\vec{\mathbf{q}}_{e},\vec{\mathbf{p}}_{t(\bar{t})}\}$} 
with given polarization $\lambda,\lambda'$. 

As in general $D^{t}$ gets also CP invariant contributions from 
absorptive parts in the SM amplitude, the truly CP violating 
contribution will be singled out through the difference: 
\be 
  {\mathcal A}^{\! {\mathcal T}}_{\lambda\lambda'}
= 
  \mbox{A}^{\! {\mathcal T}}_{b}{\scriptstyle (\lambda,\lambda')} 
- \mbox{A}^{\! {\mathcal T}}_{\bar{b}}
  {\scriptstyle (-\lambda',-\lambda)} 
\label{Observable:T} 
\ee 
where $\mbox{A}^{\!{\mathcal T}}_{\bar{b}}
{\scriptstyle (-\lambda',-\lambda)}$ 
refers to process (\ref{2}). A non--zero value of 
(\ref{Observable:T}) would imply CP violation in the 
$t\bar{t} \gamma$ and/or $t\bar{t}Z$ vertices. From (\ref{d4sigma}) 
we obtain: 
\be 
  {\mathcal A}^{\! {\mathcal T}}_{\lambda\lambda'}
= 
- \alpha_{b} \frac{3 \beta \pi \sqrt{s}}{2 m_{t}} 
  \frac{\eRe D_{1}}{N_{tot}}
\; .
\ee
It is also possible to define a triple product asymmetry for 
determining $\eRe D_{2}$. It is 
necessary to consider not only the space above and below the 
production plane as in eq.(\ref{Observable:T}), but also in addition 
the forward and backward region with respect to the direction of the 
top quark. 

The asymmetry ${\mathcal A}^{\! {\mathcal T}}_{\lambda\lambda'}$
is different from zero if CP invariance is violated in
(\ref{1}) and (\ref{2}). We have obtained the analytic expression
Eq. (\ref{Observable:T}) for 
${\mathcal A}^{\! {\mathcal T}}_{\lambda\lambda'}$ 
assuming CP violation in the production process only. However, the 
same expression will hold if CP violation occurs also in the decay 
vertex, i.e. ${\mathcal A}^{\! {\mathcal T}}_{\lambda\lambda'}$
is insensitive to CP violation in $t\to bW$. In order to
measure CP violation in $t\to bW$ through triple product
correlations one has to consider the three--body decay 
$t\to b\ell\nu$~\cite{{ECMF:1},{we}}. As
${\mathcal A}^{\! {\mathcal T}}_{\lambda\lambda'}$ is odd
under time reversal, the real parts of $\dgz$
($\eRe D_{i}$) enter in (\ref{Observable:T}). 
${\mathcal A}^{\! {\mathcal T}}_{\lambda\lambda'}$ violates
both P and CP invariance for $\lambda = \lambda' = 0$. 
Accordingly, the analytic expression obtained is proportional to 
$D_{1}$, which is a P--odd and CP--odd combination of the coupling 
constants.

\paragraph{4.} The CP violating angular asymmetries as defined above 
determine $H_{i}$ and $D_{i}$ and thus depend on the beam 
polarization. The beam polarization can strongly enhance 
(or decrease) the effects we are interested in. Measurements 
performed with opposite beam polarizations can be used to 
disentangle $H_{i}^{0}$ from $D_{i}^{0}$. 
We define the following polarization asymmetries analogous to the 
standard forward--backward asymmetry~(\ref{Asymmetry:FB}):
\be
P_{FB}^{b(\bar{b})}{\scriptstyle (\theta_{0})} 
= \frac{( 1 - \lambda \lambda' )}{( \lambda - \lambda') }
  \cdot
  \frac{(\sigma^{b(\bar{b})}_{F}
        -\sigma^{b(\bar{b})}_{B})
         {\scriptstyle (\theta_{0},\lambda,\lambda')}
       -(\sigma^{b(\bar{b})}_{F}
        -\sigma^{b(\bar{b})}_{B})
         {\scriptstyle (\theta_{0},-\lambda,-\lambda')}}
       {(\sigma^{b(\bar{b})}_{F}
        +\sigma^{b(\bar{b})}_{B})
         {\scriptstyle (\theta_{0},\lambda,\lambda')}
       +(\sigma^{b(\bar{b})}_{F}
        +\sigma^{b(\bar{b})}_{B})
         {\scriptstyle (\theta_{0},-\lambda,-\lambda')}}
\; .
\ee
Then the CP violating asymmetry is
\be 
  {\mathcal P}^{FB}_{}{\scriptstyle (\theta_{0})} 
= P_{FB}^{b}{\scriptstyle (\theta_{0})} 
+ P_{FB}^{\bar{b}}{\scriptstyle (\theta_{0})} 
\; . 
\ee
Again, keeping only the terms linear in in \mbox{$\mIm \dgz$}, we 
obtain: 
\be 
  {\mathcal P}^{FB}_{}{\scriptstyle (\theta_{0})} 
= - 6 \alpha_{b} b \cos\theta_{0} 
\left(
  2 
  \cdot
  \frac{\mIm D_{2}^{0}}
       {{\mathcal N}_{tot}^{0}{\scriptstyle (\theta_{0})}} 
- 3 \sin^{2}\theta_{0} 
  \cdot
  \frac{a_{1\lambda}^{SM}}
       {{\mathcal N}_{tot}^{0}{\scriptstyle (\theta_{0})}} 
  \cdot
  \frac{\mIm H_{1}^{0}}
       {{\mathcal N}_{tot}^{0}{\scriptstyle (\theta_{0})}} 
\right)
\; ,
\label{Polarization:FB}
\ee
where
\be
  {\mathcal N}_{tot}^{0}{\scriptstyle (\theta_{0})}
&=&
  {\mathcal N}_{tot}{\scriptstyle (\theta_{0},\lambda=\lambda'=0)}
\nn 
  a_{1\lambda}^{SM}
&=& 
  2 b G_{3}^{0}
- \alpha_{b} \left( 
      ( 1 + \beta^{2} - 2 b ) F_{1}^{0} 
    + ( 1 - \beta^{2} ) F_{2}^{0} 
  \right) 
\; .
\ee
\bigskip
We define a ``central'' polarization 
asymmetry ${\mathcal P}^{C}{\scriptstyle (\eta)}$
\be
{\mathcal P}^{C}{\scriptstyle (\eta)} 
= \frac{( 1 - \lambda \lambda' )}{( \lambda - \lambda' )}
  \frac{(\sigma^{b}_{C}-\sigma^{\bar{b}}_{C})
        {\scriptstyle (\eta,\lambda,\lambda')} 
       -(\sigma^{b}_{C}-\sigma^{\bar{b}}_{C})
        {\scriptstyle (\eta,-\lambda,-\lambda')}}
       {(\sigma^{b}_{C}+\sigma^{\bar{b}}_{C})
        {\scriptstyle (\eta,\lambda,\lambda')} 
       +(\sigma^{b}_{C}+\sigma^{\bar{b}}_{C})
        {\scriptstyle (\eta,-\lambda,-\lambda')}}
\ee
in order to measure $D_{1}^{0}$: 
\be
{\mathcal P}^{C}{\scriptstyle (\eta)} 
=
\frac{- 6 \alpha_{b} b \sin^{2}\eta \, \mIm D_{1}^{0}}
     { N_{tot}^{0}
     - \sin^{2}\eta \, 
       [ ( 3 b - 2 \beta^{2} ) F_{1}^{0}
       + 3 \alpha_{b} ( b - \beta^{2} ) G_{3}^{0} ]
     } 
\; ,
\label{PC}
\ee
where $N_{tot}^{0}=N_{tot}{\scriptscriptstyle (\lambda=\lambda'=0)}$.

\section{Numerical results within MSSM} 
So far our formulae are general and model independent. Now we want to 
give numerical results for the 
{CP violating} observables eqs.(\ref{Observable:A}), 
(\ref{Observable:FB}), (\ref{Observable:C}), 
(\ref{Observable:Z}), and (\ref{Observable:T}), defined in the 
previous section. We use the results for the electroweak dipole 
moment form factors $\dgs$ and $\dzs$ as obtained in~\cite{dipole}. 
There $\dgs$ and $\dzs$ were calculated within the MSSM~\cite{Kane} 
where we allowed for complex 
parameters. We included gluino, chargino, and neutralino exchange in 
the loop of the \mbox{$\gamma t\bar{t}$} and \mbox{$Z t\bar{t}$} 
vertex. 

The observable quantities 
${\mathcal A}^{FB}_{\lambda\lambda'}{\scriptstyle (\theta_{0})}$, 
${\mathcal A}^{A}_{\lambda\lambda'}{\scriptstyle (\theta_{0})}$, 
${\mathcal A}^{Z}_{\lambda\lambda'}{\scriptstyle (\eta)}$, 
${\mathcal A}^{C}_{\lambda\lambda'}{\scriptstyle (\eta)}$, and 
${\mathcal A}^{\! {\mathcal T}}_{\lambda\lambda'}$ 
depend on $\dgs$ and $\dzs$ and 
therefore on the parameters $M^{\prime}$, $M$, and $m_{\tilde{g}}$, 
the mass parameters of the gauge groups $U(1)$, $SU(2)$, and $SU(3)$, 
respectively, $|\mu|$, the higgsino mass parameter, 
$\tan\beta = v_{2}/v_{1}$, with $v_{i}$ being the real vacuum 
expectation values of the Higgs fields, $m_{\tilde{t}_{k}}$, 
$m_{\tilde{b}_{k}}$, the masses of the two stops and the two 
sbottoms, $\theta_{\tilde{t}}$, $\theta_{\tilde{b}}$, their mixing 
angles, $\varphi_{\tilde{t}}$, $\varphi_{\tilde{b}}$, their mixing 
phases, and $\varphi_{\mu}$, the phase of the higgsino mass 
parameter. We use the GUT relations
$m_{\tilde{g}} = (\alpha_{s}/\alpha_{2}) M \approx 3 M$ and
$M^{\prime} = \sfrac{5}{3} \tan^{2}\Theta_{\scriptscriptstyle W} M$, 
and take \mbox{$m_{\scriptscriptstyle W} = 80$~GeV}, 
\mbox{$m_{t} = 175$~GeV}, \mbox{$m_{b} = 5$~GeV}, 
\mbox{$\sqrt{s} = 500$~GeV}, 
\mbox{$\alpha_{s}{\scriptstyle (\sqrt{s})} = 0.1$}, and 
\mbox{$\alpha_{em}{\scriptstyle (\sqrt{s})} = \sfrac{1}{123}$}.
For the parameters we choose:
\begin{center}
\renewcommand{\baselinestretch}{1.} 
\renewcommand{\arraystretch}{1.} 
\begin{tabular}{|rcrl|rcrl|rcrl|}
  \hline
  $M$ &=& 230 & \hspace{-4mm} GeV &
    $m_{\tilde{t}_{1}}$ &=& 150 & \hspace{-4mm} GeV &
      $m_{\tilde{b}_{1}}$ &=& 270 & \hspace{-4mm} GeV \\
  $|\mu|$ &=&      250 & \hspace{-4mm} GeV &
    $m_{\tilde{t}_{2}}$ &=& 400 & \hspace{-4mm} GeV &
      $m_{\tilde{b}_{2}}$ &=& 280 & \hspace{-4mm} GeV \\
  $\tan\beta$ &=& 3 & &
    $\theta_{\tilde{t}}$ &=& $\frac{\pi}{9}$ & &
      $\theta_{\tilde{b}}$ &=& $\frac{\pi}{36}$ & \\
  $\varphi_{\mu}$ &=& $\frac{4 \pi}{3}$ & &
    $\varphi_{\tilde{t}}$ &=& $\frac{\pi}{6}$ & &
      $\varphi_{\tilde{b}}$ &=& $\frac{\pi}{3}$ & \\
  \hline
\end{tabular}
\renewcommand{\baselinestretch}{1.} 
\renewcommand{\arraystretch}{1.} 
\end{center}
Furthermore, the asymmetries depend on the polarizations of the 
electron and positron beams $(\lambda,\lambda')$, on the polar 
angle $\theta_{0}$, (see eq.(\ref{Observable:A})) describing the 
forward--backward cut, and the polar angle $\eta$ 
(see eqs.(\ref{Observable:Z}), (\ref{Observable:C})). For these 
quantities we take: 
\begin{center}
\renewcommand{\baselinestretch}{1.} 
\renewcommand{\arraystretch}{1.} 
\begin{tabular}{|rcrl|rcrl|rcrl|}
  \hline
  $\lambda = - \lambda'$ &=& $-0.8$, $0$, $0.8$ & &
    $\theta_{0}$ &=& $\frac{\pi}{12}$ & &
      $\cos\eta$ &=& $\frac{1}{\sqrt{3}}$ & \\
  \hline
\end{tabular}
\renewcommand{\baselinestretch}{1.} 
\renewcommand{\arraystretch}{1.} 
\end{center}
For $\lambda = -1$ the electron is purely left--handed. In all 
figures we show the asymmetries for 
$\lambda = - \lambda' = \{0 , -0.8 , 0.8 \}$ 
(full line, dashed line, dotted line). 

In Fig.~1a we show 
${\mathcal A}^{FB}_{\lambda\lambda'}{\scriptstyle (\theta_{0})}$ as a 
function of $\sqrt{s}$ for our parameter set and different 
polarizations of the electron--beam. The curves exhibit spikes due to 
thresholds of intermediary particles in the $t\bar{t}$--production. 
The spikes are already present in the dipole moment form factors 
$\mIm\dgs$ and $\mIm\dzs$, as discussed in detail in~\cite{dipole}. 

Notice that the size of the asymmetry strongly depends on the 
polarization of the electrons. For $\sqrt{s} \lsim 700$~GeV it is
much bigger (of order $0.2 \mal 10^{-3}$) if the electrons are left 
polarized. Quite generally 
${\mathcal A}^{FB}_{\lambda\lambda'}{\scriptstyle (\theta_{0})} \gsim 
2 {\mathcal A}^{A}_{\lambda\lambda'}{\scriptstyle (\theta_{0})}$ 
for $\sqrt{s} \gsim 500$~GeV. 

The dependence of 
${\mathcal A}^{Z}_{\lambda\lambda'}{\scriptstyle (\eta)}$ on 
$\sqrt{s}$, shown in Fig.~2a, is very similar to that of 
${\mathcal A}^{FB}_{\lambda\lambda'}{\scriptstyle (\theta_{0})}$ 
just discussed. It is, however, roughly $1.5$ times bigger than 
${\mathcal A}^{FB}_{\lambda\lambda'}{\scriptstyle (\theta_{0})}$ 
and two times as big as 
${\mathcal A}^{C}_{\lambda\lambda'}{\scriptstyle (\eta)}$. 
For the value $\cos\eta = \frac{1}{\sqrt{3}}$ 
${\mathcal A}^{Z}_{\lambda\lambda'}{\scriptstyle (\eta)}$ 
is maximal. 

In Fig.~1b we show the dependence of the asymmetry 
${\mathcal A}^{FB}_{\lambda\lambda'}{\scriptstyle (\theta_{0})}$ 
on the mass parameter $M$. 
Notice the interesting difference between left--handed and 
right--handed electrons in the lower mass region $M \lsim 360$~GeV. 
${\mathcal A}^{Z}_{\lambda\lambda'}{\scriptstyle (\eta)}$ again 
shows the same behaviour being $1.5$ times bigger. The decrease of 
${\mathcal A}^{FB}_{\lambda\lambda'}{\scriptstyle (\theta_{0})}$ and
${\mathcal A}^{Z}_{\lambda\lambda'}{\scriptstyle (\eta)}$ 
for bigger $M$ can be explained by the fact that also the 
dipole moment form factors $\mIm\dgs$ and $\mIm\dzs$ decrease 
because the gaugino--higgsino mixing of the charginos and 
neutralinos becomes weaker. 

The difference between left--handed and right--handed electrons is 
again apparent in the $\varphi_{\mu}$ dependence of 
${\mathcal A}^{FB}_{\lambda\lambda'}{\scriptstyle (\theta_{0})}$ as 
shown in Fig.~1c. In the case of left--handed electrons the asymmetry 
shows a $\sin\varphi_{\mu}$ behaviour, whereas in the case of 
right--handed electrons practically no $\varphi_{\mu}$ dependence 
is seen. When the CP violating phase $\varphi_{\mu}$ vanishes, one 
has still an asymmetry due to the phase $\varphi_{\tilde{t}}$ in 
the mass matrix of the $\tilde{t}$ squarks. 

In Fig.~2b one sees the different contributions to 
${\mathcal A}^{Z}_{\lambda\lambda'}{\scriptstyle (\eta)}$. 
The gluino is dominating only for small $M \lsim 130$~GeV. 
For $M \gsim 130$~GeV the charginos give the main contribution, 
and for $M \gsim 230$~GeV the neutralinos are more important than 
the gluino. This analysis of the contributing diagrams holds for all 
asymmetries, 
${\mathcal A}^{FB}_{\lambda\lambda'}{\scriptstyle (\theta_{0})}$, 
${\mathcal A}^{A}_{\lambda\lambda'}{\scriptstyle (\theta_{0})}$,
${\mathcal A}^{Z}_{\lambda\lambda'}{\scriptstyle (\eta)}$, and
${\mathcal A}^{C}_{\lambda\lambda'}{\scriptstyle (\eta)}$,
that depend on the imaginary parts of the dipole moment form factors. 
For the triple--product asymmetry 
${\mathcal A}^{\! {\mathcal T}}_{\lambda\lambda'}$ 
the situation is a bit different. Although the gluino and the 
chargino contributions are quite big, they have opposite signs and 
therefore the neutralino contribution plays a more important role 
than in the other asymmetries. 

The dependence of 
${\mathcal A}^{Z}_{\lambda\lambda'}{\scriptstyle (\eta)}$ 
on $\tan\beta$ is shown in Fig.~2c. An interesting 
fact is the very weak dependence on the polarization for large 
$\tan\beta$. There is a strong decrease of the asymmetry in 
$2 \lsim \tan\beta \lsim 10$. On the other hand the triple product 
asymmetry ${\mathcal A}^{\! {\mathcal T}}_{\lambda\lambda'}$ 
has nearly no dependence on $\tan\beta$ (not shown here).

In Fig.~3a we show the dependece of 
${\mathcal A}^{\! {\mathcal T}}_{\lambda\lambda'}$ on $\sqrt{s}$. 
The similiarity between this plot and the plots for 
$\eRe\dgs$ and $\eRe\dzs$ in~\cite{dipole} can be clearly seen. 
The spikes at $\sqrt{s}=400$~GeV and $\sqrt{s}=590$~GeV are due to 
the thresholds of $\tilde{\chi}^{+}_{1}\tilde{\chi}^{-}_{1}$ 
production with $m_{\tilde{\chi}^{+}_{1}} = 200$~GeV and 
$\tilde{\chi}^{+}_{2}\tilde{\chi}^{-}_{2}$ production with 
$m_{\tilde{\chi}^{+}_{2}} = 295$~GeV respectively. 

Fig.~3b shows the dependence of 
${\mathcal A}^{\! {\mathcal T}}_{\lambda\lambda'}$ on $M$. 
Again the behaviour reflects that of the dipole moment form factors 
$\eRe \dgs$ and $\eRe \dzs$. There is a big difference between left 
and right polarized electrons. 

The dependence of the triple--product asymmetry on the phase 
$\varphi_{\mu}$, as shown in Fig.~3c, is weaker than for the other 
asymemtries. 

The asymmetries 
${\mathcal A}^{FB}_{\lambda\lambda'}{\scriptstyle (\theta_{0})}$, 
${\mathcal A}^{A}_{\lambda\lambda'}{\scriptstyle (\theta_{0})}$, 
${\mathcal A}^{Z}_{\lambda\lambda'}{\scriptstyle (\eta)}$, and
${\mathcal A}^{C}_{\lambda\lambda'}{\scriptstyle (\eta)}$ have 
all the same shape due to the similiarity of the imaginary parts of 
the dipole moment form factors. The difference of 
${\mathcal A}^{Z}_{\lambda\lambda'}{\scriptstyle (\eta)}$ and
${\mathcal A}^{C}_{\lambda\lambda'}{\scriptstyle (\eta)}$ in the 
dependence of $\eta$ comes from the fact, that the denominator in 
${\mathcal A}^{C}_{\lambda\lambda'}{\scriptstyle (\eta)}$ decreases 
proportionally to $\cos\eta$ and therefore this factor cancels. 
Although ${\mathcal A}^{C}_{\lambda\lambda'}{\scriptstyle (\eta)}$ 
reaches its maximum for $\cos\eta=0$, there is no parameter space 
left for the measurement. Thus the best value for $\eta$ should be 
$\cos\eta = \frac{1}{\sqrt{3}}$, the maximum value for 
${\mathcal A}^{Z}_{\lambda\lambda'}{\scriptstyle (\eta)}$.

\section{Summary and Conclusions} 
The process $e^{+}e^{-} \to t\bar{t}$, with $t$($\bar{t}$)
decaying into $W^{+} b$($W^{-}\bar{b}$), is well suited to study CP 
violating effects beyond the Standard Model. We have derived 
analytic formulae for the cross section and the angular distributions 
of the $b$ and $\bar{b}$ quarks assuming CP violation in the 
\mbox{$\gamma t\bar{t}$} and \mbox{$Z t\bar{t}$} vertices. 
Particular attention was paid to the 
$t$($\bar{t}$) polarization. We then defined appropriate independent 
CP violating asymmetries, which quite generally allow one to 
determine both the real and imaginary parts of the electroweak dipole 
moment form factors of the top, $\dgs$ and $\dzs$. It was possible to 
integrate over the whole phase space and to obtain rather simple 
analytic expressions for these asymmetries. Thus far our study was 
quite general. We also performed a numerical analysis of these 
asymmetries within the Minimal Supersymmetric Standard Model with 
complex parameters. In this model the asymmetries turn out to be 
of order $\lsim 10^{-3}$.

\section*{Acknowledgements}
E.C.'s work has been supported by the Bulgarian National Science 
Foundation, Grant Ph--510. 
This work was also supported by the 'Fonds zur F\"orderung der 
wissenschaftlichen Forschung' of Austria, project no. P10843--PHY. 


\clearpage
\section*{Figure Captions}

\paragraph{Figure~1:} 
The asymmetry 
${\mathcal A}^{FB}_{\lambda\lambda'}{\scriptstyle (\theta_{0})}$ 
eq.(\ref{Observable:FB}) for the reference parameter set 
and a longitudinal electron beam 
polarization of 
-80\% (dashed line), 
  0\% (full line), and
 80\% (dotted line) depending on
\mbox{\textbf{(a)} $\sqrt{s}$~[GeV]}, 
\mbox{\textbf{(b)} $M$~[GeV]}, and
\mbox{\textbf{(c)} $\varphi_{\mu}$}. 

\paragraph{Figure~2:} 
The asymmetry 
${\mathcal A}^{Z}_{\lambda\lambda'}{\scriptstyle (\eta)}$ 
eq.(\ref{Observable:Z}) for the reference parameter set. 
\\ \textbf{(a)} Depending on \mbox{$\sqrt{s}$~[GeV]} 
for a longitudinal electron beam polarization of 
-80\% (dashed line), 
  0\% (full line), and
 80\% (dotted line). 
\\ \textbf{(b)} Depending on \mbox{$M$~[GeV]} 
for a longitudinal electron beam polarization of 
-80\% (thick dashed line) together with the different contributions: 
chargino   (thin dashed line), 
neutralino (thin dotted line), and 
gluino     (thin dashed--dotted line).
\\ \textbf{(c)} Depending on \mbox{$\varphi_{\mu}$} 
for a longitudinal electron beam polarization of 
-80\% (dashed line), 
  0\% (full line), and
 80\% (dotted line). 

\paragraph{Figure~3:} 
The asymmetry ${\mathcal A}^{\! {\mathcal T}}_{\lambda\lambda'}$ 
eq.(\ref{Observable:T}) for the reference parameter set 
and a longitudinal electron beam 
polarization of 
-80\% (dashed line), 
  0\% (full line), and
 80\% (dotted line) depending on
\mbox{\textbf{(a)} $\sqrt{s}$~[GeV]}, 
\mbox{\textbf{(b)} $M$~[GeV]}, and
\mbox{\textbf{(c)} $\varphi_{\mu}$}. 

\clearpage 

\begin{center}
\setlength{\unitlength}{1mm}
%
\begin{picture}(120,215)(0,0)
\rotatebox{90}{\put(0,0){\mbox{\epsfig{file=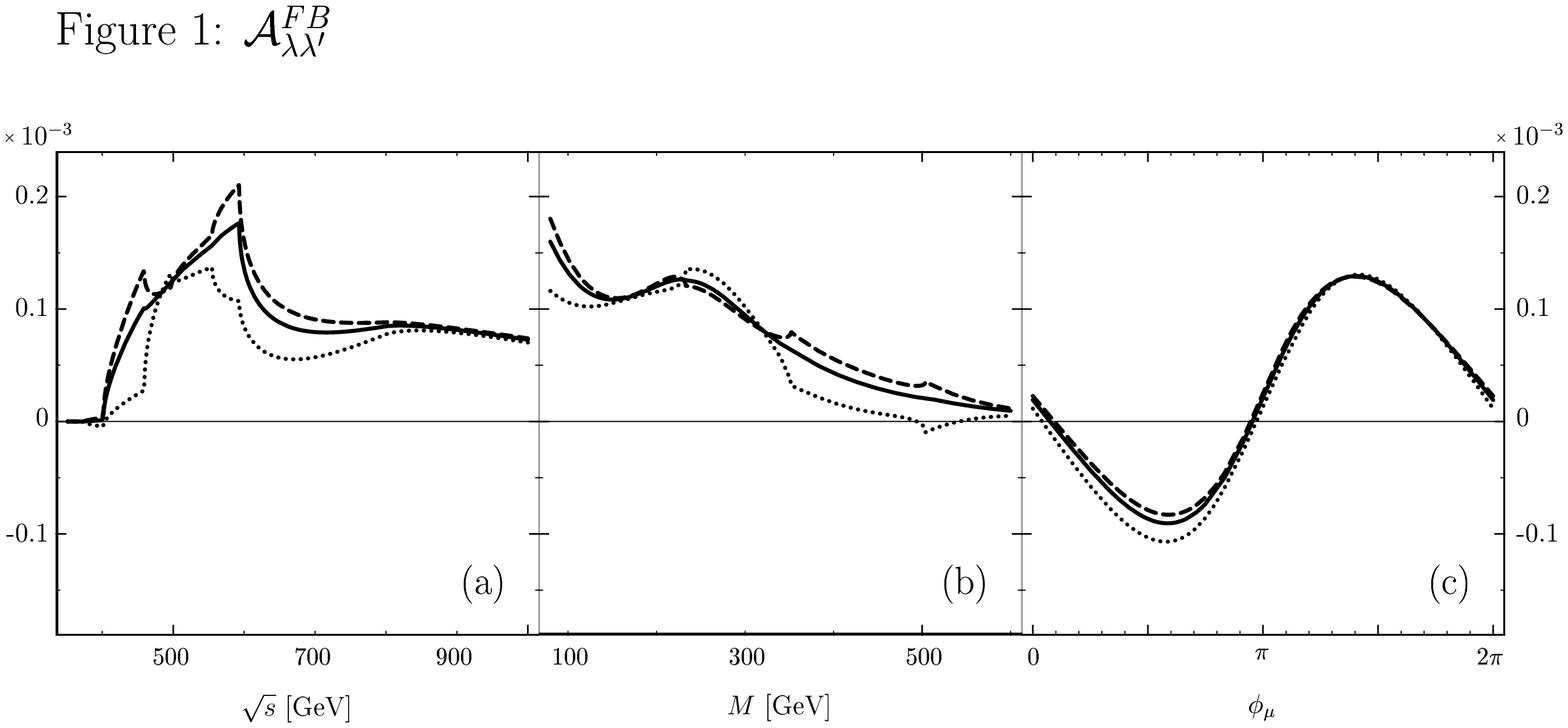,width=215mm}}}}
\end{picture}
\\
\clearpage
%
\begin{picture}(120,215)(0,0)
\rotatebox{90}{\put(0,0){\mbox{\epsfig{file=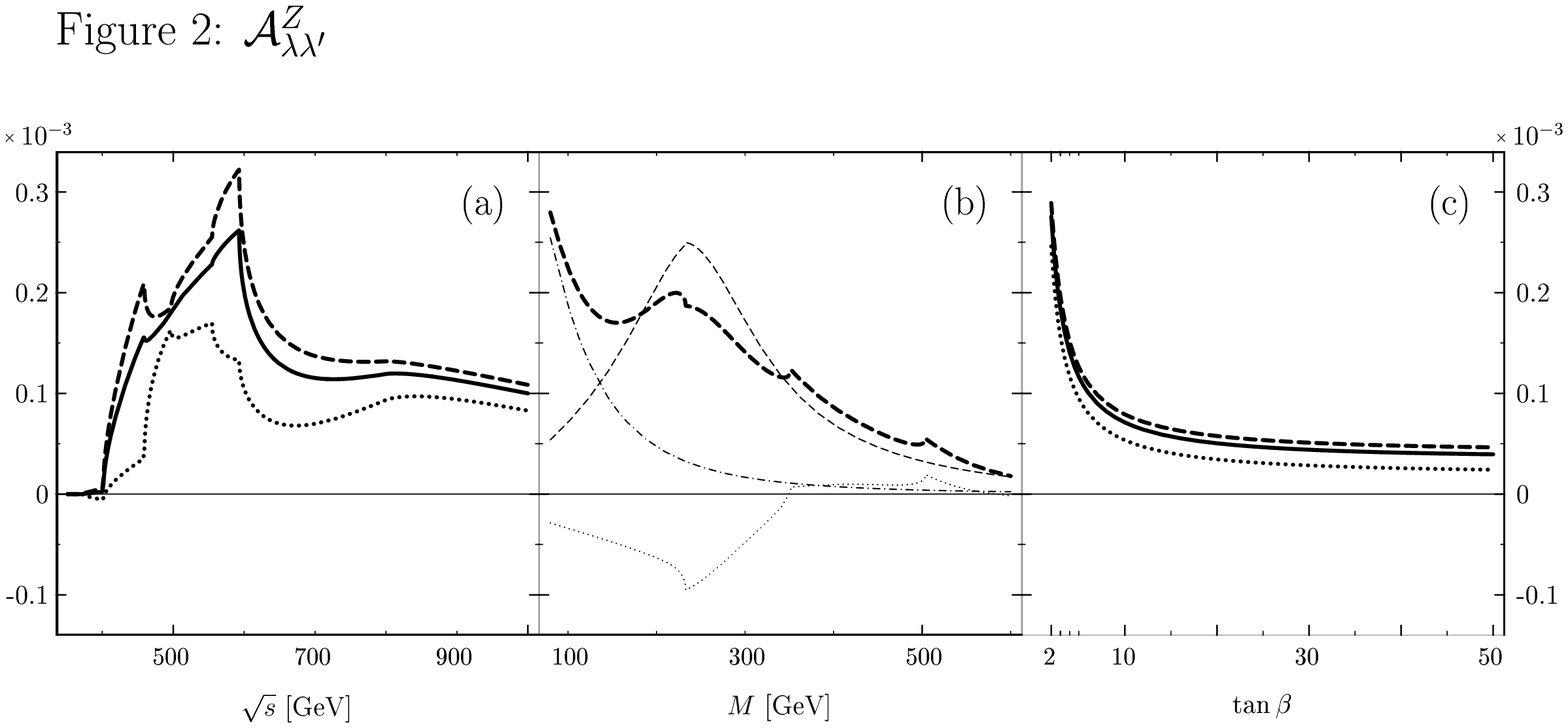,width=215mm}}}}
\end{picture}
\\
\clearpage
%
\begin{picture}(120,215)(0,0)
\rotatebox{90}{\put(0,0){\mbox{\epsfig{file=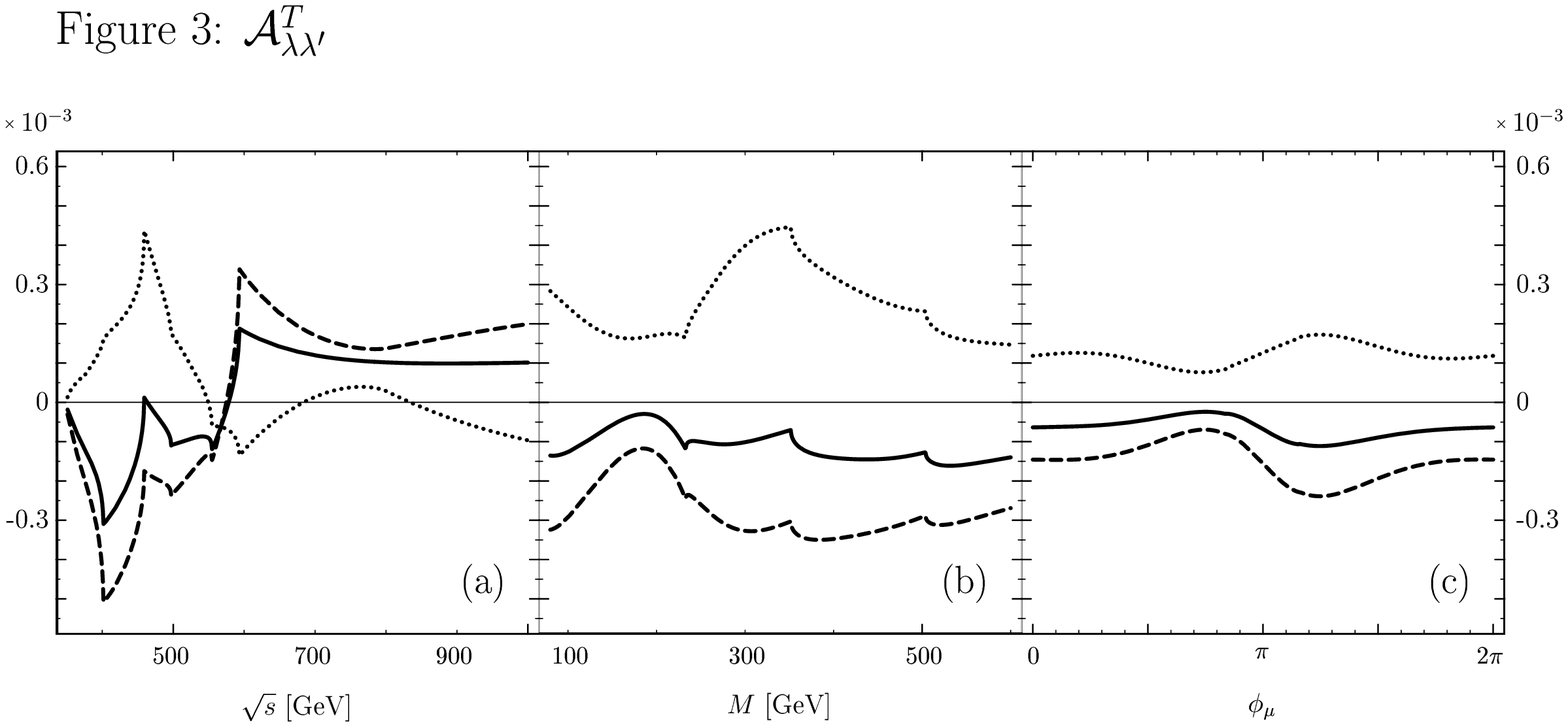,width=215mm}}}}
\end{picture}

\end{center}

\end{document}